\documentclass[aps,prl,twocolumn,superscriptaddress,groupedaddress,showpacs]{revtex4-1}  

\usepackage{amsmath,amssymb}
\usepackage{amsfonts}
\usepackage{relsize}
\usepackage{simpler-wick}
\usepackage{simplewick}
\usepackage{cancel}
\usepackage{mathtools}
\usepackage[T1]{fontenc}
\usepackage{geometry}
\usepackage{fancyhdr}
\usepackage{hyperref}

\usepackage{bibentry}

\usepackage{color}

\usepackage{graphicx}
\usepackage{dcolumn}
\usepackage{bm}
\usepackage{enumerate}
\usepackage{slashed}
\usepackage{simplewick}
\usepackage{comment}

\usepackage{soul}

\usepackage[english]{babel}
\usepackage{appendix}
\usepackage[utf8]{inputenc}

\usepackage{marginnote}
\usepackage[normalem]{ulem}

\usepackage{subfigure}

\newcommand*\xbar[1]{%
   \hbox{%
     \vbox{%
       \hrule height 0.5pt 
       \kern 1.0ex
       \hbox{%
         \kern-0.1em
         \ensuremath{#1}%
         \kern-0.1em
       }%
     }%
   }%
} 
\newcommand*\xxbar[1]{%
   \hbox{%
     \vbox{%
       \hrule height 0.5pt 
       \kern 0.8 ex
       \hbox{%
         \kern-0.1em
         \ensuremath{#1}%
         \kern-0.1em
       }%
     }%
   }%
}

\newcommand{\bz}{\bar z}

\newcommand{\ii}{{\rm{i}}}

\newcommand{\nn}{\nonumber}

\newcommand{\eq}[1]{(\ref{#1})}
\renewcommand{\>}{\rangle}
\newcommand{\la}{\label}
\newcommand{\ba}{\begin{align}}
\newcommand{\ee}{\end{equation}}
\newcommand{\be}{\begin{equation}}

\def\12{\frac{1}{2}}
\newcommand{\p}{\partial}
\newcommand{\en}{\end{align}}

\newcommand{\<}{\langle}

\usepackage{color}

\begin{document}

\setcounter{secnumdepth}{-1}

\title{ Inner  Nonlinear Waves and Inelastic Light Scattering of Fractional Quantum Hall States as  Evidence of the  Gravitational Anomaly  }

\author{P. Wiegmann}
 \affiliation{Kadanoff Center for Theoretical Physics, University of Chicago, 5640 South Ellis Avenue, Chicago, Illinois 60637, USA}

\date{\today}

\begin{abstract}
We develop the  quantum hydrodynamics of inner waves in the bulk of fractional quantum Hall states. We  show that  the   inelastic light scattering by inner waves is  a sole effect of the gravitational anomaly. 
 We  obtain   the
  formula for the oscillator-strength, or mean energy of optical absorption  expressed solely in terms of an  independently measurable static structure factor. The formula does not explicitly depend on a model interaction potential.\end{abstract}

\date{\today}

\maketitle


\newpage

\paragraph{ Introduction}Excitations  in the bulk of fractional
quantum Hall states (FQH) are neutral collective modes of density modulations.
These modes are generally  gapped.  Evidence
of  collective modes    were   seen in inelastic light
scattering
\cite{pinczuk1993,*pinczuk2001,*pinczuk2005} and  in optical absorption by surface acoustic waves   \cite{kukushkin}.
The  numerically obtained spectrum  of  a  small system \cite{Haldane_1985} and \cite{Platzman_1996},
also shows a dispersive branch 
of a collective
excitation.

The experimental accessibility  of the  dispersion of neutral   modes of  FQH states      calls for a better
 understanding
 of  inner waves. There is  a renewed interest in the subject. Some recent papers are collected in Ref. \cite{jolicoeur2017shape,*yang2012model,*maciejko2013field,*golkar2016higher,*Gromov-Son}.
 
 In this paper we  show that  inelastic light  scattering  by inner FQH waves is  a sole effect of the gravitational anomaly.  This observation  gives a  geometric
 interpretation to   inner waves, and also, 
a new analytic formula for the "oscillation strength" of optical absorption \(\Delta_k\).

 The gravitational anomaly only recently entered the quantum Hall effect (QHE) literature. Some   papers on the subject are  in Ref. \cite{CLW2014,*Abanov2014,*framinganomaly,*Klevtsov2014,
*Read2015,*KW2015,*lcw,*Klevtsov:2016_lectures}. The gravitational anomaly is an elusive phenomenon which appeared  as a higher-order gradient correction to bulk transport coefficients \cite{footnote1}.    What would be the  clean, experimentally accessible  effects of the gravitational anomaly? We argue that   the gravitational anomaly governs one of the major observables in FQH, the {\it inelastic  light scattering}.

A natural approach to studying inner  waves   is   hydrodynamics.
It   goes back to the seminal paper  \cite{girvin1986} by Girvin,
MacDonald and Platzman (GMP).  Our analysis is based on  more recent  development  of FQH hydrodynamics \cite{wiegmann2013hydrodynamics,*wiegmann2013JETP}  (see, also \cite{stone1990superfluid}). As the GMP theory, the recent hydrodynamics approach  has  roots in a similarity
between  FQH states and  a superfluid, but with the essential addition that the superfluid
is {\it rotating  and incompressible}.

We briefly describe the central point of the paper.
 The correspondence \cite{wiegmann2013hydrodynamics,*wiegmann2013JETP} between the  FQHE and a superfluid identifies electrons  
 with   quantized vortices in a    fast rotating incompressible
 superfluid. 
Hydrodynamics of a 2D incompressible
fluid can be reformulated  as {the Helmholtz law} (see, e.g., \cite{Helmholtz}):
  {\it Vortices 
are frozen (or  passively dragged
by)  the flow}. Since vortices represent electrons they could be probed by
light. Then, the Helmholtz
law forbids  inelastic light scattering.  Being perturbed by light, vortices instantaneously
change the flow and  remain frozen into a new flow. {  They cannot accelerate against the flow}.

{  Our main observation} is   that the quantization subtly corrects the Helmholtz law through the  gravitational anomaly. The inelastic light  scattering    is the effect of
 this correction. 

The  gravitational anomaly comes to the stage to prevent a quantization scheme
from violation diffeomorphism invariance, the relabeling symmetry of the fluid.  It is quite remarkable,
 that optical probes  directly test    this fundamental
symmetry. 

{  The hydrodynamic description of inner FQHE waves   faces a long-standing problem of  the quantizing of incompressible   hydrodynamics, specifically the  chiral flows, that are flows with an extensive vorticity.  Accounting  for the gravitational anomaly described below represents perhaps the first consistent quantization of incompressible flows, whose applications go beyond the QHE.}

Before we proceed, an important comment about the spectrum of incompressible waves is in order.
The GMP theory  \cite{girvin1986} adopted  a variational approach   initially developed
by Feynman for the superfluid helium \cite{feynman1954,*FEYNMAN1955}. The GMP approach  assumes
that a certain two-body Hamiltonian \(H=\sum_q V_q\rho_q\rho_{-q}\), where
\(\rho_q\) is the electronic density mode, indeed, delivers a FQH state. 
  Then it assumes that  excitations include  a single-mode density   modulation  \(|k\>=\rho_k|0\>\),
and interprets the diagonal matrix element of the Hamiltonian   \begin{align}\Delta_k=\frac{\<k|H|k\>}{\<k|k\>}.\la{12171}\end{align}
as a variational approximation to the excitation spectrum. 
The net result is expressed  in terms
of a model potential \(V_q\).

Such an approach is justified for compressible fluids, like helium,
where   atomic density modulation,  is a linear wave.   In this case a single-mode  \(|k\>=\rho_k|0\>\) is a long lived state.  Contrary to GMP's major assumption, a single-mode state 
{\it does not}   approximate a long-lived excitation  in  incompressible fluids, such as the FQHE. A reason for it is  that incompressible  waves  are essentially nonlinear. A single-mode state  decays into multiple modes, and does not have a spectrum, and    \(\Delta_k\) has no  direct relation to true excitations, as it seems commonly accepted in the literature.

Still, we argue that   \(\Delta_k\) could be measured in optical  absorption  and give  a new formula for  \(\Delta_k\)  in terms of the structure factor. It  refines the GMP formula which expresses \(\Delta_k\)  in terms of model potential $V_q$.

We focus  on  Laughlin's series of states,
where a filling
fraction  \(1/\nu\) is an integer.

 \paragraph{ Correspondence between FQH states and fast rotating  superfluid}

The   analogy  between  FQHE and a superfluid was
suggested in Refs. \cite{girvin1986,stone1990superfluid} and developed to a correspondence  in Ref.  \cite{wiegmann2013hydrodynamics,wiegmann2013JETP}. We briefly review it. 
  In short, a drift of vortices in a fast rotating superfluid  and   a motion
of electrons in FQH regime  are governed by  same
equations.

Fast rotating  superfluid  is a dense media of same sense vortices with a
quantized circulation, which we denote by  \(2\pi\Gamma\).
The total vorticity of the fluid is compensated by a solid rotation with
a  frequency  \(\Omega\),  such that the mean density of vortices  is \(\rho_0=\Omega/(\pi\Gamma)\).
 We assume that the vortices are in a liquid phase (do not crystallize).
The frequency of rotation \(\Omega\) corresponds
to the Larmor frequency   \( \Omega=eB/2m_*\) with an effective mass \(m_*\). The   "mass" is the only phenomenological parameter of the theory  determined by the spectral gap. Its   energy scale is  the Coulomb interaction \(2\hbar\Omega=\hbar^2/{ 2m_*\ell^2}
\sim e^2/ \ell\), where \(\ell=\sqrt{\hbar/eB}\) is the magnetic
length.  Then vortices correspond   to
 electrons  if the vortex circulation in units of  \({\hbar}/{  m_*}\)  is the inverse
of the  filling fraction and the gap in the spectrum is of the order of $\hbar\Omega$
\begin{align}
\Gamma=\left(\frac{\hbar}{  m_*}\right)\nu^{-1},\quad \Omega= \frac{eB}{2m_*}
.\la{0} 
\end{align}
 
  This correspondence  differs from that of GMP  \cite{girvin1986}.
  The authors of Ref. \cite{girvin1986} referred to the  work of Feynman  \cite{feynman1954}, who considered   atomic
density modes of a compressible  superfluid at rest. Rather, we discuss the
modes of vorticity of a rotating incompressible superfluid  \cite{footnote2}.
 We will measure the distance in units of magnetic length  and the energy  (the bulk gap) in units of the $2\hbar\Omega$, setting
  \(\ell=\hbar=m_*=1\). In these units, the mean density \(\rho_0=1/(2\pi\Gamma)={\nu}/{2\pi}\).
We restore the parameters in final formulas.
The filling fraction \(\nu\) plays a role of a semiclassical parameter.

\paragraph{ Helmholtz law} The hydrodynamics of a 2D incompressible flow can
be cast in a compact    Helmholtz  form: {\it\ The material derivative of vorticity
vanishes}. If \(\bm u=(u_x,u_y)\) is the velocity of  a flow,   \(\omega={\bm\nabla}\times{\bm
u}\) is the vorticity,   
\(D_t=\p_t+\bm u\cdot\bm\nabla\) is the  material derivative, and the fluid
is incompressible
\({\bm \nabla}\cdot{\bm u}=0\), then the Euler equation in the Helmholtz
form  reads
\begin{align}
D_t\omega=0.\la{21}
\end{align}

In the context of FQH vorticity
is identified with the electronic density. In a rotating frame with no  net vorticity the correspondence between electronic density and vorticity is    \begin{align}
\rho({\bm r})=\rho_0+\frac
 1{2\pi\Gamma}\omega ({\bm r}).\la{31}
\end{align}   The velocity of the flow \(\bm u\) does not have a  measurable
analog in the  FQHE. It   could be thought as a transversal part of the  fictitious
gauge field attaching a  flux of magnetic field to electrons.

It is quite remarkable that  essential features of  Laughlin's  states
are  encapsulated  in the  quantum version of the Helmholtz equation.
We will see some   of it now.

The Helmholtz law reflects a geometric meaning of  hydrodynamics:
Incompressible flows are generated by a successive action of volume-preserving diffeomorphisms. In the QHE this concept has been  suggested
in Ref.  \cite{cappelli1993}. Therefore, the dynamics of FQH inner waves, and the equivalent  problem of a quantum   hydrodynamics, both are  seen as a more abstract problem of  the quantization
of the group of volume-preserving diffeomorphisms.
This group  is generated by density mode operators \(\bar\rho_k = \int
e^{-\ii{\bm k}\cdot{\bm r}} \bar\rho({\bm r})\  d^2{\bm r}\),  with the algebra 
\begin{align}
[\bar\rho_k,\ \bar\rho_{ k'}]=\ii e_{kk'}\bar\rho_{ k+ k'},
\la{4}
\end{align}   
{with the structure constants (in the long-wave approximation)   
\begin{align*}e_{kk'}={\bm k}\times{\bm k'}.\end{align*}
On  the torus the structure constants  are $e_{kk'}=2 e^{1/2(\bm
k\cdot\bm k')}\sin(\tfrac 12\bm k\times{\bm
k}')$.  Here   we used {\it  bar} to emphasize the quantization as in }Ref.\cite{girvin1986}.  The  classical limit of \eq{4} is the  Poisson brackets of  hydrodynamics  (see e.g., \cite{Marsden_1983}). 


\paragraph{ Nonlinear waves } Few important properties already follow from  \eq{21}. One is that inner density waves  are essentially
nonlinear. A well-known fact is that the 2D incompressible hydrodynamics does not assume  linear waves.  In the language of the quantum
theory, this means that single density modes  are not  long-lived states.     

However, the Euler equation can be linearized about an inhomogeneous background. Example
 are Tkachenko linear modes of a vortex crystal \cite{
tkachenko1966stability}.
 Also, if we impose a periodic density modulation \(|k_0\>\), then on top
of it there
are linear waves \(\bar \rho_{q-k_0}|k_0\>=\bar\rho_{q-k_0}\bar \rho_{k_0}|0\>\). 

This suggests that, in contrast to a single mode, the  two-modes states do have a spectrum. This assertion agrees with the interpretation  of  the inelastic light scattering experiments of Pinczuk {\it\ et all}
\cite{pinczuk1993,*pinczuk2001,*pinczuk2005} as a Raman type two-modes process by  Platzman and He \cite{Platzman_1996}. 
In the  case of two-modes excitations the mean energy
\(\Delta_{q,k_0}=\frac{\<q,k_0|H|q,k_0\>}{\<q,k_0|q,k_0\>}\) is true variational  approximation
of the spectrum. In the limit \(k_0\to  0\) the   waves become nonlinear,
and the variational approach fails. 
We address the spectrum of inner waves elsewhere.

Another consequence   mentioned already is that the Helmholtz law prohibits  the absorption
of  light.  The central point of this paper is to show how this problem is   resolved by the  quantization. 
\paragraph{ Quantization of Euler equation }
Quantization of the Euler equation meets essential difficulties.
The advection term ${\bm u}\cdot\nabla\omega=\nabla({\bm u\cdot} \omega)$ where 
 two operators  sit at the same point requires   a regularization.
The problem in a general setting has  a long history of failures
 and commonly considered nearly impossible. A  scheme of  regularization where   points are split ${\bm u}({\bm r}+\frac\epsilon
2) \omega({\bm r}-\tfrac\epsilon 2)$ leads to inconsistencies. The difficulty is that the point- splitting distance itself depends
on the flow $\epsilon[{\bm u}]$.  Hence, a regularization scheme is
specific to the flow and cannot be  practical  to all  varieties of
flows at once.  However, in  a special case,
when  the flow  consists  of  a dense media of   vortices,  quantization
 can be achieved.  In this case, a variable   short-distance
cutoff is the distance between  vortices  \(\epsilon\sim 1/\sqrt{ \rho}\).
 Below  we present a  heuristic, but an economic
approach to quantization.

We will use the complex notations. We denote the complex velocity
by \(u_z=u_x-\ii u_y\) and  use the stream function \(\psi\) and the traceless part of the fluid momentum flux tensor \(\Pi_{ij}=u_iu_j-(1/2)\delta_{ij}{\bm u}^2\). In   complex coordinates \(u_z=2\ii\p_z\psi
\) and  
 \(\Pi_{zz}=
u_z^2\). We will write the advection term  as \begin{align}
{\bm u}\cdot{\bm \nabla}\omega&={\ii[\p_{z}^2\Pi_{\bar z\bz}-\p_{\bz}^2\Pi_{zz}].}\la{53}
\end{align} 
 Hence, we have to understand
the quantum meaning of
 \begin{align}
\Pi_{zz}\equiv u_z^2=-4(\p_z\psi)^2,\la{52}
\end{align}
 For that we recall a notion of the projected density operator. 
\paragraph{  Normal ordering and quantization} 
States on the  lowest Landau level (LLL), and also flows of rotating superfluid, are realized as Bargmann space \cite{girvin1986,bargmann1961}. It is a space of holomorphic
functions with the inner product \(\<g|f\>=\int e^{-1/2|z|^2}\ g^\ast( \bar z)f(z)
dz d\bz \).  The density operators acting in the Bargmann space obeying the algebra \eq{4} are realized by  the normally ordered  operator     \begin{align}\bar\rho_k=\sum_ie^{-\frac{\ii}
2{\rm
k}z_i^\dag}e^{-\frac{\ii} 2{\rm
\bar k}z_i}, 
 \end{align}
 where \({\rm k}=k_x+\ii k_y\) is a complex wave vector and  \begin{align}
z_i^\dag=2\p_{z_i}.\nn
\end{align}   In  \cite{girvin1986} it was   called a projected (onto LLL) density operator. It is organized such  that states   \( |k\>=\bar \rho_{
k}|0\>\)  is holomorphic and, hence belongs to LLL.  It is
also chiral  \(\overline\rho_k^\dag=\bar\rho_{-k}\). 
The projected density mode  operator is obtained from the density mode \(\rho_k=\sum_ie^{-\ii{\bm
k\cdot\bm r}_i}\)    by  positioning the anti-holomorphic coordinates to  the left to holomorphic coordinates, and  replacing them by the holomorphic  differentiating  operators \(\bar z_i\to 2\p_{z_i} \). As a result   matrix elements between
states on LLL (with respect to Bargmann inner product) are the same as that of density modes. Similarly, the   two-modes operator that  entered the momentum flux tensor on the Bargmann space is represented by a normal ordered string
 \begin{align}\overline{\rho_k\rho_{k'}}=\sum_{i,j}
e^{-\frac\ii 2{\rm
k}z_i^\dag}e^{-\frac\ii 2{\rm
k'}z_j^\dag}e^{-\frac \ii 2{\rm
k}^*z_i}e^{-\frac \ii 2{\rm
k'}^*z_j}.\la{78}\end{align} 
The projected density modes generate coherent states
 of  LLL and also the states of  rotating superfluid if \(z_i\) is the coordinate of a vortex.

We denote the Wick contraction  \(\wick[offset=0.8em]{\c A\c B}=\overline {AB}-\bar
A\bar B\) and compute  \(\wick[offset=0.6em]{\c u_z\c u_z}\). The  Wick contraction of    two   density modes follows from \eq{78}
 \begin{align}
\wick[offset=0.6em]{\c \rho_k\c \rho_{k'}}=\bar \rho_{k+k'}(1-e^{\12{{\bm k}\cdot {\bm k'}}}).\la{81}
\end{align}

The next step  is to express the momentum flux tensor \eq{52} (acting  on the Bargmann space) in terms of   the generators  \(\overline\rho_k\). 
We get  insight by computing it for the ground state where the density is uniform $\rho_k=N\delta_{k0}$ and there is no flow.

Equation \eq{81} gives   the  contraction of two stream functions. At the uniform state it reads 
\begin{align}
\wick[offset=0.9em]{\c\psi({\bm r})\c \psi({\bm r}')}=
 \frac{2\pi}{\nu} \int e^{\ii{\bm
k}\cdot ({\bm r}-{\bm r}')}\Big(\frac{1-e^{-\12{k^2}}}{k^4}\Big)\frac{d^2k}{(2\pi)^2}.\la{85}
\end{align}
Now we can 
 compute  \(\wick[offset=0.7em]{\c u_z({\bm r})\c u_z({\bm r'})}=-4\p_z\p_{z'}\wick[offset=0.9em]{\c\psi({\bm r})\c \psi({\bm r}')}\). In the hydrodynamic limit 
 \be |r - r'|\gg \ell:\quad   \wick[offset=0.7em]{\c u_z({\bm
r})\c u_z({\bm r'})}\sim (z-z')^{-2}.\ee
 As \(r\to r'\),
the divergency  cutoff  by the magnetic length, but the  net result  is zero anyway in the state with the  rotational symmetry. The effect of short-distance divergency does not show up.  It is nulled by averaging over the angle.

\paragraph{ Gravitational anomaly in hydrodynamics}
Now  we evaluate  \(u^2\)    on a   flow state where the  density and  the cutoff \(\epsilon[\bm u]\) at \(r\to r'\) are  not uniform.  The result  follows from
the geometric  interpretation of the fluid flow.  In this picture   the distance
between particles (vortices)     is interpreted  as a metric \(ds^2=\rho|
dz|^2\) of a auxiliary Riemann surface, and a flow as an evolving surface. The (scalar)
curvature of the auxiliary surface is
\begin{align}
\mathcal{R}=-4\rho^{-1}\p_z\p_{\bz}\log\rho.\la{10} 
\end{align} 

The distance between particles  is invariant under a change
of coordinates, which can be seen as a relabeling of fluid particles.
 In hydrodynamics,
this fictitious  symmetry is typically  applied
to fluid atoms. In our approach, it is a relabeling symmetry of vortices.
We want to keep this major symmetry intact in  quantization.

To proceed, we first notice that in the hydrodynamic limit the contraction of stream functions
\eq{85} is the Green function of the Laplace operator \begin{align}
 \wick{\c\psi({\bm r})\c \psi({\bm r}')}=
 \frac\pi\nu G({\bm r},{\bm r}').
\end{align}
It is natural to assume that in a flow state the contraction is the Green function of the Laplace-Beltrami operator in the metric $\rho |dz|^2$. The Green   function diverges at \(r\to r'\) and requires a regularization. There is   only one
covariant regularization. It    identifies the short- distance cutoff with the
geodesic distance \(d({\bm
r}, {\bm r}')\).
With this prescription, we  define the Wick contraction of the momentum flux
tensor \eq{52} as a limit:  \begin{align}
\wick[offset=0.65em]{\c u_z\c u_z}= \frac{4\pi}\nu \lim_{r\to r'}\partial_{z}
\partial_{z'}\Big[G({\bm
r}, {\bm r}')+\frac 1{2\pi}\log d({\bm
r}, {\bm r}')\Big].\la{92}
\end{align} The result  of this limit  is  well known (Supplemented Material): The short distance expansion
of \eq{92}  is the Schwarzian of the metric \begin{align}
\wick[offset=0.65em]{\c u_z\c u_z}= 
\frac{1}{6\nu} \left( \partial_z^{2}\log\rho-
\frac{1}{2}(\partial _z\log \rho)^{2}\right).\nn
\end{align}
Then the Wick contraction of the advection term \eq{53} is expressed through the
curvature \eq{10}  of the auxiliary Riemann surface\begin{align}
\wick[offset=0.7em]{\c {\bm u}\cdot{\bm\nabla}\c \omega}=\frac{1}{ 96\pi}
\ {\bm\nabla}\mathcal{R\times {\bm\nabla}\omega}.\la{121}
\end{align} 
 This is the main result of the quantization \cite{footnote3}. We can now treat the hydrodynamics  as a field theory, with a   constant
  cutoff, independent of the flow.  The cutoff is the  magnetic length. The price for this is that  the   material derivative is no longer
zero. With
the help of  \eq{121} we obtain
 \begin{align}
D_t\bar \rho=\frac{1}{ 96\pi}\mathcal{R}\times{\bm\nabla}\bar\rho.\la{62}
\end{align}

If waves are small,   
\(\mathcal{ R}\approx-\rho_0^{-2}\Delta\rho\), the  correction to the
Helmholtz law  could  be
treated in the harmonic approximation \begin{align}
 D_t\bar\rho_k=
  \frac \pi{24\nu^2}
  \sum_q q^2(\bm k\times\bm q)\bar\rho_{q}\bar\rho_{k-q}.\la{63}
\end{align}
We emphasize that Eq.\eq{62}  incurs higher order gradients and powers of
curvature, both suppressed by a small magnetic length.
Later we show how to generate higher gradient corrections in the harmonic approximation.

  \paragraph{ Deviation from  Helmholtz law} The implication of quantum corrections
 is that the Helmholtz law held  for quantum operators  does not
hold for their matrix elements: The material derivative for the projected density
mode \eq{62} does not vanish. Acceleration of particles against the flow
appears in higher derivatives,
 and it is a quantum correction, 
but, as we will see,  it    is  the only source for the light scattering. 

One can ask whether   model specific effects beyond the  hydrodynamics could
change the coefficient $1/24$ in \eq{63}. We do not see such possibility.
The coefficient has  a topological origin and  is quantized in units of 
$1/24$
or  $1/48$ depending on  a  FQHE state.

The universal departure from  the  Helmholtz law   is the central result of the paper.

\paragraph{ Hamiltonian and current}

Now we are  in a position to  determine  the Hamiltonian which  yields Eq.
\eq{62}. We write the Helmholtz equation as  a continuity
equation  
 \begin{align}
\dot{\rho}+\bm\nabla\cdot {\bm J}=0.\la{32}
\end{align} and cast it in the Hamiltonian \(\dot\rho=\frac{\ii}\hbar[H,\rho]\).
Using the commutation relation for two functionals of density 
$$[F, G]=\frac
\nu{2\pi\ii}\int\left[{\bm\nabla}
\frac{\delta F}{\delta \rho}\times {\bm\nabla}\frac{\delta G}{\delta \rho}\right]\rho\,
dV$$
 followed form  \eq{4}, we obtain  the formula for the  current   \begin{align}
 \bm { J}\bm =\frac  1{2\pi\Gamma}\rho\ \bm\nabla\times\frac{\delta   H}{\delta\rho}.\la{141}
\end{align}
Then we determine the current from \eq{62} and compute the Hamiltonian.  We write the formulas in a semiclassical manner treating $\rho$ as the  mean density of the flow (not an operator).  tThe
result for the current is\begin{align}
 \bm { J}=\rho\left( {\bm u}+
 \frac 1{96\pi}
 {\bm\nabla}\times\mathcal{R}\right)-\frac \Gamma 4\left(1 -2 \nu\right){\bm\nabla}\times\rho.\la{17}
\end{align}
The last term here is the  divergent-free part  of the current which does
not enter the equation \eq{32} and is determined separately. We comment
on its origin later.

From there we determine the Hamiltonian (in the Eulerian specification). We write it by separating  classical
and quantum contributions
and restoring the units\begin{align}
&H=\int\left(\mathcal{H}-\hbar S\right) \rho_0 d^2{\bm r},\la{151}\\
&{\mathcal H}=\frac{m_*}2\left(
 {\bm u}^2+2\bm u\cdot\bm u_0-\pi\Gamma^2 \rho\log \rho \right),\la{54}\\
&S=-\pi\Gamma\left( 
\rho\log \rho  +
\frac{1}{96\pi} (\nabla \log \rho)^2
\right).\la{74}
\end{align}
For references we write the Hamiltonian  in the  form which separates overall scale and emphasize \(\nu\) as a semiclassical parameter\begin{align}
H=\frac{\hbar^2}{2\nu m_*\ell^2}&\int \Big[2\pi \rho({\bm r})G({\bm r},{\bm r'})\rho({\bm r'})d^2{\bm r}'+\nn \\
& 
(\nu-\frac{1}{2})\rho\log \rho  +
\frac{\nu}{96\pi} (\nabla \log \rho)^2
 \Big] d^2{\bm r.}\nn
\end{align}  Here \(G\) is the
Green function of the Laplace operator.

The first two  terms  in the classical part \eq{54} are the kinetic  and
 centrifugal energies, \(\nabla\times{\bm u}_0=2\Omega\).  The last term
 in \eq{54}  regularizes
the divergency  of the kinetic energy at vortex
cores. It was known in the theory of superfluid 
since the  1961 paper of Kemoklidze and Khalatnikov \citep{kemoklidze,*Khalatnikov},
see also more recent Ref. \citep{AW}. This term is the
Casimir invariant, whose
 Poisson bracket  with all other local fields vanishes. For this reason,  does
not show in Eqs.(\ref{62},\ref{32}).
It enters the current as a  divergence-free term.

The quantum part \eq{74} at a fixed vortex circulation \(\Gamma\) does not
depend on \(m_*\). It also  consists of two terms. The first term is a quantum
correction to the  Kemoklidze-Khalatnikov term. 


The second term in \eq{74} (also the second term in \eq{17} and the
RHS of \eq{62}) is the effect
of the gravitational anomaly.

\paragraph{ Static structure factor} 
Now we show that the hydrodynamic equations  encapsulate independently known
 long wave  expansion of the   structure factor.  This
 fact could serve as a check and  justification of the hydrodynamic equations (\ref{62}-\ref{74}).

According to the general theory of linear response, the structure factor  $$s_k=\tfrac
1N\<0|{\rho_{-k}\rho_k}|0\>=\tfrac 1N\<0|\overline{\rho_{-k}\rho_k}|0\>,$$
appears  in the  harmonic   approximation of the current, or as a  the rigidity
of density modes in the  Hamiltonian (see Supplemented Material)  
\begin{align}\la{270}
&{\bm J}_k\approx\tfrac 1 N \sum_{q\neq 0}\ii{\bm q}^*s^{-1}_q\overline{\rho_{k-q}\rho_q},\\ & H\approx 
\tfrac 1{2N}\sum_{q\neq 0} s_q^{-1} \overline{\rho_{-q}\rho_q}.\la{27}
\end{align}
 Here \({\bm q}^*\) is the vector normal to \(\bm q\), and   the  ground state energy is set to zero. 

We compute the inverse structure factor by expanding \eq{151}.
The result is \begin{align}
s^{-1}_q= \frac{2}{q^2 } -\left(\frac 1{2\nu}  -1 \right)  +[s^{-1}_q]_{_\mathsmaller{+}},\la{222}
\end{align} where \([s^{-1}_q]_+\) is the  part of the expansion which consists of positive powers of the momentum.
The leading term in \([s^{-1}_q]_{_\mathsmaller{+}}\) is the  effect of the gravitational
anomaly
 \begin{align}
  [s^{-1}_q]_{_\mathsmaller{+}} = \frac{q^2}{24\nu} +\mathcal{O}(q^4).\la{22}
  \end{align}

  Inverting \eq{222}, we obtain
the first three terms of the small $q$  expansion of the structure factor
 \begin{align}
   s_q =  \frac{q^2}{2} + \frac {q^4}{8\nu} (1 - 2\nu )  + \frac{q^6}{8\nu^2}(\tfrac
34 - \nu)(\tfrac 13 - \nu) \la{23}
 \end{align}
Each of the three terms in these formulas is independently known,has a
universal meaning,  and reflects
   symmetries of the hydrodynamics and Laughlin states. The term \(q^2\)
in
\eq{23} corresponds to the kinetic energy of the fluid, \(\12{\bm u}^2\),
and referred as the `perfect screening'
 sum rule;
\(q^4 \)  corresponds to the \(\rho\log\rho\) term in (\ref{54},\ref{74})
and is referred as the "compressibility" sum rule. Finally,  the \(q^6\) term represents
the gravitational anomaly. It  was first  obtained in Ref.\citep{Kalinay2000}
directly from the Laughlin wave function. In  equivalent forms,   it  appeared
 independently in Ref.\citep{CLWBig}.

Using  \eq{23} and \eq{81} we obtain the long-wave  expansion of the projected
structure factor  $$\bar s_k=\tfrac
1N\<0|\bar \rho_{-k}\bar \rho_k|0\>. $$
From \eq{81} we have  \(\bar s_q=s_q-(1-e^{-\12 q^2})\). Hence, 
\begin{align}
\bar s_q=(1-\nu )\frac {q^4}{8\nu}  \left(1+\frac 1{6\nu}(3-10\nu)q^2\right)+\dots\la{25}
 \end{align}
 There is no apparent reasons to think that higher terms
in the expansion, but the first three, are universal.

\paragraph{ Harmonic approximation  }We can now express the  
correction to the Helmholtz law in terms of  the   structure factor, refining
Eq.  \eq{63}. 
Let us  compute $[H,\bar\rho_k]$ with the Hamiltonian \eq{27}. The first term in the expansion of    $s_q^{-1}
$  \eq{222} forms the material derivative, and  the second does not contribute.  
The correction to the Helmholtz law is due to the positive part of
the expansion \eq{222}, whose leading term is the gravitational anomaly \eq{22}.
We obtain a refine form of  the Eq.\eq{63} valid at all \(k:\)
\begin{align}
 D_t\bar\rho_k=\frac{\pi}{\nu } \sum_q  e_{kq}[s^{-1}_q]_{_\mathsmaller{+}}\bar\rho_{q}\bar\rho_{k-q}.\la{28}
\end{align}

At \(k\to 0\)  \eq{28} reduces to \eq{63}.

We comment, that the harmonic approximation does not mean that   waves are
 linear,
or that the state \(\bar\rho_k|0\>\) is an excitation, as it seems suggested
in \citep{girvin1986}.

\paragraph{ Optical absorption by nonlinear  waves} 
Now we are ready to compute the  optical absorption.
Absorption occurs  when  light accelerates particles
against the flow, i.e., due to a departure  from the  Helmholtz law. 

Consider an acoustic wave   imposed through the Hall bar as in an experiment  
\cite{kukushkin}. It creates  a state \( |k\>=\overline\rho_k|0\>\).
In solids, the optical absorption measures the differential  intensity 
\( S_k(\omega)=\frac 1N\<k|\delta(H-\hbar\omega)|k\>\) and the
  integrated intensity \( \bar s_k=\hbar\int S_k(\omega) d\omega=\frac1N\<k|k\>\),
the projected static structure factor. 
Another object of interest in spectroscopy is  the oscillation strength, the first moment of the intensity
  \begin{align}  \bar f_k = \int  \omega S_k(\omega)d\omega =  \frac
1N \<k|H|k\> =  \frac
{\ii}{2N} \<0|\dot{\bar\rho}_k\ \bar\rho_{-k}|0\>
\la{281}
\end{align}
 and the   the  mean energy \eq{12171}  
 \be\Delta_k={
\bar f_k}/{\bar s_k}={\<k|H|k\>}/{\<k|k\>}.\la{me}\ee

In  fluids,  a  proper definition of  the intensity must be written in a
coordinate system moving with the fluid.  This means that  the    time derivative
in \eq{281} is the material  derivative 
 \begin{align}
\bar f_k=&\frac {1} {2N\ii}\<0|[D_t\bar\rho_{k},\ \bar\rho_{-k}]|0\>\la{91}.
\end{align} 
   Hence, only the rhs of \eq{28} enters \eq{91}. 

Typically \(S_k(\omega)\)  features an asymmetric peak supported
by the curve \(\hbar \omega=\Delta_k\),
which is, rudimentarily interpreted as a spectrum of excitations. Such an interpretation will be valid, would   \(\bar
\rho_{
k}|0\>\) be a long-lived  state, as
  happens in a compressible fluid. As we commented above, in the  FQHE, the state \(\bar
\rho_{k}|0\>\) is  one of many short-lived coherent states.

Interpretation aside, we  compute \(\bar f_k\). Equation \eq{28}  reduces \eq{91}  to 
\(\<0|\bar\rho_{-k}\bar\rho_{q}\bar\rho_{k-q}|0\>\), which we compute with the
help of the algebra (\ref{4}) and  the structure constants \eq{4} for the torus.  We  express result for the mean energy \eq{me}  in terms  of 
$$ \tilde s_k= (1-\nu)^{-1}e^{\12
{k^2}}\bar s_k\nn.$$
In units  
 \(\hbar^2/(\pi m_*\ell^2)={ 2}\hbar\Omega/\pi\) the mean energy reads 
  \begin{equation}\la{30}
\scalebox{0.95}[1]{$  \Delta_k=\tilde s_k^{-1}\int   \sin^2(\12\bm k\times\bm
q)e^{-\frac{q^2}2}[s^{-1}_q]_{_\mathsmaller{+}}(\tilde s_q-\tilde s_{k-q})d^2q$}.
\end{equation}
Contrary to Eq. (4.15) of Ref. \cite{girvin1986}, our formula  does not explicitly depend on a model interaction. It is expressed only through independently measured  (or calculated) structure factor. We emphasize that beyond  terms in \eq{25} the structure factor depends on details of the material, an so as the mean energy \eq{30}.
\smallskip
 \paragraph{Magnetoroton minimum}
Both \(\bar f_k\) and \(\bar s_k\)  feature a broad 
asymmetric peak at \(k\ell\sim 1\), both vanish at \(k=0,\infty\), but their
ratio,  \(\Delta_k\), is finite and non-zero. 

 The limiting values of the structure factor are 
\begin{align}\nn
&k \to  0:\quad \ \tilde s_k\sim
\tfrac {k^4}{8\nu},\quad \; [s^{-1}_k]_{_\mathsmaller{+}}\sim\tfrac{k^2}{24\nu},\\
&
k\to \infty:\quad  
\tilde s_k= 1,\qquad [s^{-1}_k]_{_\mathsmaller{+}}=\tfrac 1{2\nu}.\nn
\end{align}
 Hence,
the mean-energy \(\Delta_k\) smoothly interpolates  between  
 \begin{align}\nn
  &\Delta_{k= 0}= 4\nu 
  \int  q^2[s^{-1}_q]_{_\mathsmaller{+}}\left(\nabla^2_q
  \tilde s_q\right)e^{-\frac{q^2}2}{d^2q}
  \end{align}
 and
  \begin{align}\Delta_{k\to \infty}= \int    [s^{-1}_q]_{_\mathsmaller{+}}( \tilde s_q+ 1) e^{ -\frac
{q^2}2}d^2q\nn
\end{align}
featuring a broad minimum at $k\ell\sim 1$.
The  minimum   also  shown in numerically evaluated   $\Delta_k$ from  model Hamiltonians \cite{girvin1986}. GMP called it magnetoroton minimum.  The minimum  is also featured by Eq. \eq{30}, however  we failed  to recognize its  universal meaning and physics behind it. It relies  on features of \(\bar s_k\) beyond its  universal part (\ref{222}-\ref{25}).

We are not aware of  an experimental
evidence of the minimum in the  absorption spectrum  of  Laughlin's states, such as  \(\nu =\frac 13\) state. A sequence of minima in optical absorption are  reported in Ref. \cite{kukushkin} for fractions other than $\frac 13$. It is not clear whether they are related to the GMP minimum for Laughlin's states. \medskip

The author thanks A. Cappelli, A.Gromov, and G. Volovik for discussions and interest to this work. The work was
supported by the NSF under  Grant NSF DMR-1206648. The author thanks the Gordon
and Betty Moore Foundation, EPiQS Initiative through Grant GBMF4302 for the
hospitality at Stanford University, and the Brazilian Ministry of Education (MEC) and the UFRN-FUNPEC for the hospitality at IIP during the work on this paper.
\onecolumngrid
\section{Supplemented Material}
\paragraph{Structure factor and harmonic approximation}Here we give a  sketch of the derivation of harmonic approximation of the  current and the Hamiltonian (\ref{270},\ref{27}). In the harmonic approximation they are
\begin{align}
{\bm J}_k\approx\tfrac {1}{N}  \sum_{q\neq 0}{\ii\bm q}^*s^{-1}_q\overline{\rho_{k-q}\rho_q},\qquad H\approx 
\tfrac{1}{2N}\sum_{q\neq 0} s_q^{-1} \overline{\rho_{-q}\rho_q}.\la{300}
\end{align}
Here \({\bm q}^*\) is the vector normal to \(\bm q\), and   the
 ground
state energy is set to zero.

Let us apply a small testing  external potential \(A_0\)
by adding a  term \(A_0\rho\) to the Hamiltonian \eq{151}. According to \eq{141}
the effect of the external potential shifts the current  by the Lorentz drift current \(\rho {\bm E^*}\), where \({\bm E}=-{\bm \nabla} A_0 \)
is the  electric field, and \( {\bm E^*}\) is the vector normal to \( {\bm E}\). Hence, in a steady
state \({\bm J}=\rho {\bm E^*}\), the Hall effect. From here we can find the density mode triggered by a non-uniform electric field. In the harmonic approximation in density mode, in the Fourier space, we write  \( {\delta\bm J}_k/\delta\rho_{k-q}= {\ii {\bm q}^*}A_{0q}\).      Further variation in \(A_0\) gives \begin{align}
 \ii {\bm q}^*=\frac{\delta^2{\bm J}_k}{\delta A_{0q}\delta\rho_{k-q}}=\frac{\delta\rho_{{q}}}{\delta
A_{0q}}\frac{\delta^2{\bm J}_k}{\delta \rho_{q}\delta\rho_{k-q}}=\<0|
{\rho_{-q}\rho_{q}}|0 \>\frac{\delta^2{\bm J}_k}{\delta \rho_q\delta\rho_{k-q}},
\end{align}where we used the   relation \(\delta\rho_{q}=\<0| {\rho_{-q}\rho_q}|0 \>\delta
A_{0q}\)  and the notation \(s_q=\frac 1N\<0| {\rho_{-q}\rho_q}|0
\>\).  The formula \eq{300} for the current follows.  The Hamiltonian follows from the relation \eq{141} \(\bm { J}\bm =(2\pi\Gamma)^{-1}\rho\ \bm\nabla\times\frac{\delta   H}{\delta\rho}\).

\bigskip

\paragraph{Short distance expansion of the Green function} Here we comment on how to compute the short distance expansion of the regularized Green function   on a Riemann surface with the metric \(ds^2=\rho |dz|^2\)  \eq{92}\begin{align}
G^R({\bm
r}, {\bm r}')=G({\bm
r}, {\bm r}')+\frac 1{2\pi}\log d({\bm
r}, {\bm r}').
\end{align}This result is standard. We included it here (and in \citep{CLWBig}), because of failure to find a proper reference.

At close points the Green   function behaves as 
\(
G({\bm
r}, {\bm r}') \to - \frac 1{2\pi}\log |{\bm
r} {-\bm r}'|.
\) At the same time the short distance expansion of the geodesic distance   reads (see, e.g., \cite{Ferrari2012})\begin{align*}
2 \log d({\bm
r}, {\bm r}') & =2  \log |{\bm
r} {-\bm r}'| +\log\rho+\frac 12(z' - z)\partial \log\rho+ \frac 12(\bar{z}' - \bar{z}) \bar{\partial} \log\rho\\ \nn
&- \frac{1}{12} |z' - z|^{2} |\partial \log\rho |^{2} + \frac{1}{48}\left[(z' - z)\partial \log\rho + (\bar{z}' - \bar{z}) \bar{\partial}\log\rho\right]^{2}\\ \nn
& + \frac{1}{6} \left[(z' - z)^{2} \partial^{2} + 2 |z' - z|^{2} \partial \bar{\partial} + (\bar{z}' - \bar{z})^{2} \bar{\partial}^{2}\right] \log\rho \,   + ...
\end{align*}
The first term of this  expansion  cancels the divergent part of the Green function.  Taking holomorphic derivatives in \(z\) and \(z'\) we obtain
\begin{align}
\lim_{r\to r'}\partial_{z}
\partial_{z'}\Big[G({\bm
r}, {\bm r}')+\frac 1{2\pi}\log d({\bm
r}, {\bm r}')\Big]=\frac{1}{24\pi} \left( \partial_z^{2}\log\rho-
\frac{1}{2}(\partial _z\log \rho)^{2}\right).\nn
\end{align}
\twocolumngrid

\begin{thebibliography}{10}
\bibitem{footnote2}{In
 the  superfluid helium  the mass \(m_*\) is the mass of a large
number of  atoms swept by a vortex, and the filling fraction \(\nu\) is number of vortices per atom.  Hence, the quantum 
corrections are suppressed by a
small vortex concentration. We
thank  G. Volovik for clarifying this aspect.
}
\bibitem{footnote3}{The regularization we described is analogous to that of Polyakov in quantizing 2D gravity}.
\end{thebibliography}

\end{document}